\documentclass[11pt]{article}
\usepackage[utf8]{inputenc}
\usepackage{epsfig}
\usepackage{amsmath,amsfonts,latexsym, amssymb}

\newtheorem{satz}{Theorem}[section]

\newtheorem{assumption}[satz]{Assumption}

\newtheorem{conclusion}[satz]{Conclusion}
\newtheorem{ob}[satz]{Observation}

\newtheorem{propo}[satz]{Proposition}

\newcommand{\mcal}{\mathcal}

\newcommand{\tit}{\textit}

\begin{document}
\thispagestyle{empty}
\begin{center}
\vspace*{1.0cm}

{\LARGE{\bf Higher-Order-Schmidt-Representations and their
    Relevance for the Basis-Ambiguity     }} 

\vskip 1.5cm

{\large {\bf Manfred Requardt }} 

\vskip 0.5 cm 

Institut f\"ur Theoretische Physik \\ 
Universit\"at G\"ottingen \\ 
Friedrich-Hund-Platz 1 \\ 
37077 G\"ottingen \quad Germany\\
(E-mail: requardt@theorie.physik.uni-goettingen.de)

\end{center}

\vspace{0.5 cm}

\begin{abstract}
 With the help of a useful mathematical tool, the polar decomposition
 of closed operators, and a simple observation, i.e. the unique relation
 between tensor-product states and compact operators, we manage to
 give a compact and coherent account of the various properties of
 higher-order-Schmidt-representations.

\end{abstract} \newpage
\setcounter{page}{1}
\section{Introduction}
In the \tit{environment-induced decoherence} approach to the quantum
measurement problem (just to mention a few sources from the huge field
of published literature, see
e.g. \cite{Zurek1},\cite{Zurek2},\cite{Joos1}), an important role is
played by the \tit{Schmidt-representation} in the tensor-product of
two Hilbert spaces, here the Hilbert space of the quantum system,
$\mcal{H}_s$, and the Hilbert space of the measuring apparatus (or
rather of the \tit{pointer}), $\mcal{H}_A$. It is argued that the
transition
\begin{equation}\left(\sum_i c_i\psi_i\right)\otimes\Phi_0\rightarrow\sum_i
    c_i\psi_i\otimes \Phi_i     \end{equation}
with
\begin{equation}\sum_i c_i\psi_i\in \mcal{H}_s\quad ,\quad
  \Phi_0,\Phi_i\in \mcal{H}_A         \end{equation}
is not yet a measurement of a quantum property of some micro object
observable but only a so-called \tit{premeasurement}.

As such this statement is quite uncontroversial as long as the pointer
system is also not of a macroscopic size. One can, on the other hand,
question the above description of the first stage of the measurement
process if the $\Phi_i$ (as is usually actually the case) belong to a
macroscopic subsystem of the measurement instrument (we will discuss
this problem elsewhere \cite{Requ1}).

It is then argued that $\sum c_i\psi_i\otimes\Phi_i$ cannot be
associated with the following mixed state of the quantum system
\begin{equation}\sum |c_i|^2|\psi_i><\psi_i|      \end{equation}
because of the well-known \tit{basis non-uniqueness} problem. To reach
a unique representation an entanglement of the pointer states with the
environment is invoked, i.e.
\begin{equation}\sum c_i\psi_i\otimes\Phi_i\rightarrow \sum c_i\psi_i\otimes\Phi_i\otimes\varepsilon_i    \end{equation}
with $\varepsilon_i$ in the ideal case an orthonormal basis of the
environment. It is then argued that this second process of correlation
makes the above representation unique.

We do not know how long this result was actually known in full
generality in the scientific community before the rigorous proofs
provided in \cite{Bub1} and \cite{Peres1}. One finds for example
sometimes statements like
\begin{equation}\sum c_i\psi_i\Phi_i=\sum c_j'\psi_j'\Phi_j'  \end{equation}
suggesting that it may happen that the sets $\{c_i\}$ and $\{c_j'\}$
could be really different (which can in fact not! happen as we will
show below). Be that as it may, we will in the following present a
very brief, mathematically concise and transparent deduction of
results being of relevance in this context. The whole line of
reasoning can essentially be based on a single observation and one
mathematical conceptual tool.
\section{The Mathematical Tool}
The mathematical tool we are employing is the very useful concept of
\tit{polar decomposition} of operators. 
\begin{satz}A closable operator from $\mcal{H}_1$ to $\mcal{H}_2$
  admits a polar decomposition of the form
\begin{equation}A=U\circ |A|       \end{equation}
which is essentially unique. $|A|$ is the positive (s.a.) operator
\begin{equation}\left(A^+A\right)^{1/2}:\mcal{H}_1\rightarrow
  \mcal{H}_1     \end{equation}
and $U$ is a partial isometry
\begin{equation}U:|A|\circ\mcal{H}_1\rightarrow A\circ\mcal{H}_1     \end{equation}
\end{satz}
Remark: As far as we know, the polar decomposition in its general form
was introduced by v.Neumann (based on earlier work of E.Schmidt,
\cite{Neumann1}). See also \cite{Kato} or \cite{Reed1}.\\[0.3cm]
\begin{satz}[Canonical Representation of a Compact Operator]
With $A:\mcal{H}_1\rightarrow\mcal{H}_2$ compact, we have
\begin{equation}A=\sum_{\nu}\lambda_{\nu}\,|\phi_{\nu}><\psi_{\nu}|    \end{equation}All
$\lambda_{\nu}\neq 0$ are only finitely degenerated and can be chosen
positive (note that phase factors can be absorbed in the ON-bases
$\phi_{\nu},\psi_{\nu}$). The possible zero-eigenspace may be infinitely
degenerated.
\end{satz}
\begin{ob}With the help of the polar decomposition we can conclude
\begin{equation}A=U\circ |A|\quad,\quad |A|\circ
  \psi_{\nu}=\lambda_{\nu}\psi_{\nu}\quad,\quad
  U\circ\psi_{\nu}=\phi_{\nu}    \end{equation}
i.e.
\begin{equation}|A|=\sum\lambda_{\nu}|\psi_{\nu}>< \psi_{\nu}|  \end{equation}
that is, the $\lambda_{\nu},\psi_{\nu}$ are the eigenvalues and
eigenvectors of $|A|$.
\end{ob} 
\section{The Schmidt-Representation}
From the above polar decomposition of a compact and, in particular,
Hilbert-Schmidt operator the Schmidt-representation follows
immediately. 
\begin{ob}A vector $\Psi$ in $\mcal{H}_1\otimes\mcal{H}_2$,
\begin{equation}\Psi=\sum c_{ij}e_i\otimes f_j  \end{equation}
$\{e_i\otimes f_j\}$ an orthonormal basis in
$\mcal{H}_1\otimes\mcal{H}_2$, can be uniquely associated with an
operator from $\mcal{H}_1\rightarrow\mcal{H}_2$ or vice versa,
i.e. with
\begin{equation}A:=\sum c_{ij}|e_i><f_j|:\mcal{H}_2\rightarrow\mcal{H}_1  \end{equation}
\end{ob}

With $\Psi$ normalisable, $A$ is Hilbert-Schmidt, hence compact. It
follows $A=U\circ |A|$ with $|A|$ having the spectral representation
\begin{equation}|A|=\sum \lambda_i|\phi_i><\phi_i|\quad,\quad \phi_i\in\mcal{H}_2  \end{equation}
and
\begin{equation}U\circ\Phi_i=\psi_i\in\mcal{H}_1  \end{equation}
thus
\begin{equation}A=\sum \lambda_i|\psi_i><\phi_i|  \end{equation}
and
\begin{equation}\Psi=\sum \lambda_i\cdot\psi_i\otimes\phi_i  \end{equation}
the latter being the Schmidt-representation.\\[0.3cm]
Remark: without mentioning it always, the eigenbasis of $|A|$
comprises also the subspace belonging to the eigenvalue zero, which,
on the other hand, does not show up in the representation of
$\Psi$.\vspace{0.3cm}

Before we proceed, we want to employ the above representation to
derive a few other results which are useful in the study of
e.g. \tit{entanglement-entropy} etc. 
\begin{ob}With
\begin{equation}A:\mcal{H}_2\rightarrow\mcal{H}_1\quad,\quad A=\sum
  c_{ij}|e_i><f_j|    \end{equation}
we have
\begin{equation}A^+=\sum \overline{c}_{ij}|f_j><e_i|    \end{equation}
and
\begin{equation}A^+A=\sum c_{ij}\overline{c}_{ij'}|f_{j'}><f_j|   \end{equation}
\begin{equation}AA^+=\sum c_{i'j}\overline{c}_{ij}|e_{i'}><e_i|   \end{equation}
\end{ob}
Furthermore we have 
\begin{ob}$A^+A$ is the reduced density matrix of $P_{\Psi}$ in $\mcal{H}_2$.
\end{ob}
Proof: We have
\begin{equation}\left(\Psi|1\otimes B|\Psi\right)=\sum
  |\lambda_i|^2\left(\phi_i|B|\phi_i\right)=\operatorname{Tr}\,\left((A^+A)B\right)   \end{equation}
\section{The Uniqueness-Question for $\mcal{H}_1\otimes\mcal{H}_2$}
If all $\lambda_i\neq 0$ are different, the spectral representation in
the above form of $|A|$ is unique apart from the possibly
degenerated zero-eigenspace. If some $\lambda_i\neq 0$ are degenerate,
for example, $\lambda:=\lambda_1=\lambda_2=\cdots =\lambda_k$, we can
choose in the $k$-dimensional subspace $\mcal{H}_{\lambda}$
arbitrarily many different $ON$-bases, connected whith each other by
unitary transformations
\begin{equation}\{\phi_1,\phi_2,\ldots ,\phi_k\}\rightarrow
  \{\phi'_1,\phi'_2,\ldots ,\phi'_k\}\quad , \quad \phi_i'=V\circ\phi_1  \end{equation}
with $V$ unitary in $\mcal{H}_{\lambda}$.

In this situation we have 
\begin{equation}A=\sum \lambda_i\cdot|\psi_i><\phi_i|=\sum
  \lambda_i\cdot|\psi'_i><\phi'_i|\quad ,\quad
  \psi_i'=U\circ\phi_i'  \end{equation}
\begin{ob}In the case of a degeneracy the $\{\psi_i,\phi_i\}$ may be
  replaced by $\{\psi_i',\phi_i'\}$ but the weights $\lambda_i$ remain
    the same. They represent the unique eigenvalues of $|A|$. That is,
    we have by the same token 
\begin{equation}\Psi=\sum\lambda_i\psi_{i,\nu}\otimes\phi_{i,\nu}=\sum\lambda_i\psi_{i,\nu}'\otimes\phi_{i,\nu}'      \end{equation}
with $\nu$ denoting the possible degeneration in the subspaces $\mcal{H}_{\lambda_i}$
\end{ob}
\section{The Schmidt-Representation for three and more Hilbert-Spaces}
Let $\Psi$ be a vector in
$\mcal{H}_1\otimes\mcal{H}_2\otimes\mcal{H}_3$, i.e.
\begin{equation}\Psi=\sum
  c_{ijk}\psi_i^1\otimes\psi_j^2\otimes\psi_k^3      \end{equation}
\begin{assumption}It exists a Schmidt-representation for $\Psi$, i.e.
\begin{equation}\Psi=\sum \lambda_i\Phi_i^1\otimes\Phi_i^2\otimes\Phi_i^3     \end{equation}
with $\lambda_i$ positive and $\{\Phi_i^{\nu}\}$ (parts of) $ON$-bases
in $\mcal{H}_{\nu}$.
\end{assumption}
\begin{ob}We associate the operator 
\begin{equation}A:\mcal{H}_1\otimes\mcal{H}_2\rightarrow\mcal{H}_3\quad
  ,\quad A:=\sum c_{ijk}|\psi_k^3><\psi_i^1\otimes\psi_j^2|     \end{equation}
with $\Psi$, so that again the $\lambda_i$ are the eigenvalues of the
operator
\begin{equation}|A|:\mcal{H}_1\otimes\mcal{H}_2\rightarrow\
  \mcal{H}_1\otimes\mcal{H}_2\quad ,\quad |A|=\sum \lambda_i|\Phi_i^1\otimes\Phi_i^2><\Phi_i^1\otimes\Phi_i^2|   \end{equation}
and $A$ can be written as
\begin{equation}A=U\circ |A|=\sum \lambda_i|\Phi_i^3><\Phi_i^1\otimes\Phi_i^2|
         \end{equation}
\end{ob}

\begin{propo}The Schmidt-representation (if it exists!) is unique even if some
  $\lambda_i$ are degenerate.
\end{propo}
Proof: We learned above that even if we assume that two different
representations do exist, i.e.
\begin{equation}\Psi=\sum \lambda_i\Phi_i^1\otimes\Phi_i^2\otimes\Phi_i^3 =
\sum \lambda'_j(\Phi'_j)^1\otimes (\Phi'_j)^2\otimes (\Phi'_j)^3  \end{equation}
the sets $\{\lambda_i\}$ and $\{\lambda'_j\}$ are necessarily the same
as both represent the unique set of eigenvalues of $|A|$. Hence it
remains only the possibility
\begin{equation}\Psi=\sum_{i,\nu} \lambda_i\Phi_{i,\nu}^1\otimes\Phi_{i,\nu}^2\otimes\Phi_{i,\nu}^3 = \sum \lambda_i(\Phi'_{i,\nu})^1\otimes (\Phi'_{i,\nu})^2\otimes (\Phi'_{i,\nu})^3      \end{equation}
where $\nu$ denote the possible degeneration of the eigenvalues
$\lambda_i$. $\Phi_{i,\nu}^1\otimes\Phi_{i,\nu}^2$ are elements in
the eigenspace of $\lambda_{i_0}$ (belonging to $|A|$) while $U$ maps them onto
$\Phi_{i,\nu}^3$. The same holds for the rhs of the equation.

Assume now that e.g. $\lambda_{i_0}>0$ is degenerate. Then we have for
the corresponding part of $\Psi$:
\begin{equation}\Psi_{\lambda_{i_0}}=\lambda_{i_0}\cdot\sum_{\nu=1}^N \Phi_{i,\nu}^1\otimes\Phi_{i,\nu}^2\otimes\Phi_{i,\nu}^3=\lambda_{i_0}\cdot\sum_{\nu=1}^N \Phi'_{i,\nu})^1\otimes (\Phi'_{i,\nu})^2\otimes\Phi'_{i,\nu})^3    \end{equation}
Note that in contrast to the two-Hilbert space case, the existence of
an eigenbase for $\lambda_{i_0}$ of the above homogeneous form is
rather special. We show that only one homogeneous eigenbase can exist
in the case of three or more Hilbert spaces.

In the eigenspace $\mcal{H}_{\lambda_{i0}}$ with the assumed two
homogeneous bases $\Phi_{i,\nu}^1\otimes\Phi_{i,\nu}^2$ and
$(\Phi'_{i,\nu})^1\otimes (\Phi'_{i,\nu})^2$ we can write
\begin{equation} \Phi_{i,\nu}^1\otimes\Phi_{i,\nu}^2=\sum_{\mu=1}^N
  c_{\nu\mu} (\Phi'_{i,\mu})^1\otimes (\Phi'_{i,\mu})^2   \end{equation}
Take now an operator $B$ in $\mcal{H}_1$ and hence $B\otimes 1$ in
$\mcal{H}_1\otimes\mcal{H}_2$. It follows
\begin{equation}\left(\Phi_{i,\nu}^1\otimes\Phi_{i,\nu}^2|B\otimes
    1|\Phi_{i,\nu}^1\otimes\Phi_{i,\nu}^2\right)=
  \left(\Phi_{i,\nu}^1|B|\Phi_{i,\nu}^1\right)= \sum_{\mu=1}^N|c_{\nu\mu}|^2 \left((\Phi'_{i,\mu})^1|B|(\Phi'_{i,\mu})^1\right)   \end{equation}
with $\sum |c_{\nu\mu}|=1$ and $N>1$ (degeneracy).
\begin{ob}As a consequence of our assumption a pure state on
  $\mcal{B}(\mcal{H}_1)$ equals a mixture. This is not! possible (see
  the appendix for a proof of this well-known result).
\end{ob}
We hence arrive at a contradiction and the proposition is proved.

Another question which was for example adressed by Peres
(\cite{Peres1}) is, how special such a homogeneous
Schmidt-representation is for more than two Hilbert spaces. A simple
counting analysis suggests, that it is in fact quite special. In our
framework we can give a complete and general answer. We exemplify the
analysis for the case of three Hilbert spaces and make the necessary
generalisations afterwards.

Take a general vector state in
$\mcal{H}_1\otimes\mcal{H}_2\otimes\mcal{H}_3$,
\begin{equation}\Psi=\sum c_{ijk}e_i\otimes f_j\otimes g_k   \end{equation}
with bases $e_i,f_j,g_k$ in
$\mcal{H}_1,\mcal{H}_2,\mcal{H}_3$. Under what conditions can $\Psi$
be represented as
\begin{equation}\Psi=\sum
  \lambda_i\cdot\Phi_i^1\otimes\Phi_i^2\otimes\Phi_i^3   \end{equation}
with (partial) bases $\Phi_i^1,\Phi_i^2,\Phi_i^3$ in $\mcal{H}_1,\mcal{H}_2,\mcal{H}_3$. 

We learned above that the possibility of such a representation is
associated with operators $A,U,|A|$ so that $\Phi_i^1\otimes\Phi_i^2$
can be extended to an eigenbasis of $|A|$ with the above
$\Phi_i^1\otimes\Phi_i^2$ belonging
to the set of non-zero eigenvalues, $A_+$, of $|A|$.\\[0.3cm]
Remark: Note that the basis vectors, belonging to $\lambda_0=0$ need
not! be of such a diagonal form.\vspace{0.3cm}

We infer from the analysis of the Schmidt-representation of the
twofold tensor product that the operator $A$, induced by $\Psi$, has
always a representation
\begin{equation}A=\sum_l |U\circ u_l><u_l|   \end{equation}
$u_l$ being the eigenvectors of
$|A|:\mcal{H}_1\otimes\mcal{H}_2\rightarrow\mcal{H}_1\otimes\mcal{H}_2$. However,
in general the $u_l$ are not of product form! The general
representation reads
\begin{equation}u_l=a^{ij}_l\cdot e_i\otimes f_j\quad\text{(summation convention)}       \end{equation}

In the following we can restrict our analysis to the subspace, $V$,
being spanned by the eigenbasis belonging to the positive
eigenvalues. If a Schmidt-representation exists, $V$ is spanned by an
eigenbasis of the form $\{\Phi_l^1\otimes\Phi_l^2\}$. For $l$ fixed we
then have
\begin{equation}\Phi_l^1=x^ie_i\quad ,\quad \Phi_l^2=y^jf_j     \end{equation}
and
\begin{equation}\Phi_l^1\otimes\Phi_l^2=x^iy^j\,e_i\otimes f_j   \end{equation}
The matrix $\left(x^iy^j\right)$ has the following form. Take e.g. a
$n\times n$ submatrix
\begin{equation}\begin{pmatrix} x^1y^1 & \cdots & x^1y^n \\ \vdots &
    \cdots & \vdots \\ x^ny^1 & \cdots & x^ny^n \end{pmatrix}          \end{equation}

All rows are proportional to each other, e.g.
\begin{equation}x^iy^1=y^1/y^2\cdot (x^iy^2)\quad\text{etc.}   \end{equation}
We hence have
\begin{ob}The rank of the matrix $\left(x^iy^j\right)$ is one.
\end{ob}
\begin{conclusion} In case of a generic $\Psi$, the respective
  eigenvectors $u_l=a_l^{ij}e_i\otimes f_j$ have matrices
  $\left(a_l^{ij}\right)$ ($l$ fixed) with a rank which is generically of
    the order $\mcal{O}(\min(\dim\mcal{H}_1,\dim\mcal{H}_2))$, i.e.,
    the probability of having such a particular Schmidt-basis is very
    small in general.
\end{conclusion}

For the tensor product of more than three spaces, we can put the
question in the following form. We ask for the probability that an
arbitrary vector can be represented as a product, that is
\begin{equation}c^{ijk\ldots n}e_i\otimes f_j\cdots\otimes
  r_n=\left(a^ie_i\right)\otimes
  \left(b^jf_j\right)\otimes\cdots\otimes \left(q^nr_n\right)        \end{equation}
For reasons of simplicity we assume that all Hilbert spaces have the
same dimension $N$. On the rhs we have $N\cdot n$ unknowns $a^i,\ldots
,q^n$. The different $c^{ijk\ldots n}$ yield $N^n$ equations.
\begin{conclusion}i) In the generic case the system of equations is
  strongly overdetermined as $N^n>N\cdot n$ for $N\geq 2$ and $n\geq
  3$.\\
ii) On the other hand, for $N=n=2$ we have $N^n=N\cdot n$. However,
this does guarantee that the system has a solution (see the following
example).
\end{conclusion}

Take e.g. the former singulett state
\begin{equation}\Psi=\left(\uparrow\downarrow-\downarrow\uparrow\right)      \end{equation}
It is entangled, hence cannot be written as a product. The above
system of equations would yield:
\begin{equation}0=a^1b^1\;,\;0=a^2b^2\;,\;1=a^1b^2\;,\;-1=a^2b^1       \end{equation}
which is contradictory. This is possible since the equations are non-linear!

\section{Appendix}
The relation 
\begin{equation}\left(\psi|B|\psi\right)=\sum_1^N \lambda_i
  \left(\phi_i|B|\phi_i\right)  \end{equation}
with $|\psi|^2=1$, $\phi_i$ being orthonormal and $0<\lambda_i<1$ for
$N>1$ because of $\sum_1^N\lambda_i=1$, which is assumed to hold for the
full algebra $\mcal{B}(\mcal{H})$, implies the identity
\begin{equation}P_{\psi}=\sum_1^N\lambda_i\cdot P_{\phi_i}  \end{equation}
We hence have
\begin{equation}1=\left(\psi|P_{\psi}|\psi\right)=\sum_1^N\lambda_i\cdot
  |\left(\psi|\phi_i\right)|^2    \end{equation}
with $0<\lambda_i<1$, $|\left(\psi|\phi_i\right)|^2\leq 1$ and
$|\left(\psi|\phi_i\right)|^2=1$ only if
$\psi=e^{i\alpha}\cdot\phi_i$. I.e., there exists at least one term
for which holds $|\left(\psi|\phi_i\right)|^2< 1$. This implies
\begin{equation}1= \sum_1^N\lambda_i\cdot
  |\left(\psi|\phi_i\right)|^2<1   \end{equation}
which is a contradiction.


\begin{thebibliography}{99}
  {\small
\bibitem{Zurek1}W.H.Zurek: ``Environment-induced superselection
  rules'', P.R. D26(1982)1862
\bibitem{Zurek2}W.H.Zurek: ``Decoherence and the Transition from
  Quantum to Classical--Revisited'', Los Alamos Science 27,2002 
\bibitem{Joos1}E.Joos,H.D.Zeh,C.Kiefer,D.Giulini,J.Kupsch,I.-O.Stamatescu:
  ``Decoherence and the Appearance of a Classical World in Quantum
  Theory'', sec.ed., Springer, Berlin 2003
\bibitem{Requ1}M.Requardt: ``An Alternative to
  Decoherence-by-Environment and the Appearance of a Classsical
  World'', preprint Goettingen 2010
\bibitem{Bub1}A.Elby,J.Bub: ``Triorthogonal uniqueness theorem and its
  relevance to the interpretation of quantum mechanics'',
  P.P. A49(1994)4213
\bibitem{Peres1}A.Peres: ``Higher order Schmidt decomposition'', PL A(1995)16
\bibitem{Neumann1}J.v.Neumann: ``Ueber adjungierte
  Funktionaloperatoren'', Ann.Math. 33(1932)294
\bibitem{Kato}T.Kato: ``Perturbation Theory of Linear Operators'',
  Springer, N.Y. 1966
\bibitem{Reed1}M.Reed,B.Simon: ``Methods of Modern Mathematical
  Physics I'', Acad.Pr., N.Y. 1980






}
\end{thebibliography}
\end{document}